\documentclass[sigconf]{acmart}

\usepackage{float}
\usepackage{makecell}
\PassOptionsToPackage{hyphens}{url}
\usepackage{hyperref}
\usepackage{listings}
\usepackage{xcolor}
\usepackage{soul}
\usepackage{microtype}
\AtBeginDocument{%
  }

\copyrightyear{2026}
\acmYear{2026}
\setcopyright{cc}
\setcctype{by}
\acmConference[ICMI '26]{INTERNATIONAL CONFERENCE ON MULTIMODAL INTERACTION}{October 05--09, 2026}{Napoli, Italy}
\acmBooktitle{INTERNATIONAL CONFERENCE ON MULTIMODAL INTERACTION (ICMI '26), October 05--09, 2026, Napoli, Italy}
\acmDOI{10.1145/3776574.3831128}
\acmISBN{979-8-4007-2318-6/2026/10}

\begin{document}

\title{TANDE: Disentangling Verbal and Nonverbal Backchannels in Emotional AI-Avatar Conversations with Young Adults}

\author{Ann-Kareen Gedeus}
\email{ag2637@cornell.edu}
\orcid{1234-5678-9012}
\affiliation{%
  \institution{Cornell University}
  \city{New York}
  \state{New York}
  \country{USA}
}
\author{Jack Good}
\email{jlg444@cornell.edu}
\orcid{1234-5678-9012}
\affiliation{%
  \institution{Cornell University}
  \city{Ithaca}
  \state{New York}
  \country{USA}
}
\author{Nadine Wagener}
\email{nadine.wagener@offis.de}
\orcid{1234-5678-9012}
\affiliation{%
  \institution{OFFIS Institute}
  \city{Oldenburg}
  \country{Germany}
}
\author{Angelique Taylor}
\email{amt298@cornell.edu}
\orcid{1234-5678-9012}
\affiliation{%
  \institution{Cornell University}
  \city{New York}
  \state{New York}
  \country{USA}
}

\renewcommand{\shortauthors}{Gedeus et al.}

\begin{abstract}
Embodied conversational agents (ECAs) need effective \textit{empathic grounding} to foster social support and engagement. Expanding into emotional domains, ECAs now use Large Language Models (LLMs) and multimodal human-agent interactions to enhance their capabilities. Yet, understanding the impact of backchanneling modalities on young adults and their gender remains limited. We introduce TANDE, an LLM-powered ECA designed for emotional conversations with young adults, a population experiencing mental, personal, and social issues with limited tools to address them. In a within-subjects study with $N=36$ young adults, we explore nonverbal and combined verbal-and-nonverbal backchanneling modalities on rapport, empathy, and engagement and isolate for gender differences. Our research shows the importance of nuanced backchanneling cues with emotional ECAs with young adults, showing a preference for nonverbal cues. We derive design implications for more effective ECAs for emotional support and well-being in young adults. The code is available at \url{https://github.com/Cornell-Tech-AIRLab/TANDE}.

\end{abstract}



\maketitle

\section{Introduction}

Embodied conversational agents (ECAs) offer a promising pathway to supplement mental well-being, with virtual agents\cite{lucas2014s} and robots \cite{jang2024minimal, kobuki2023robotic} demonstrating a greater desire in people to disclose sensitive information openly with fewer feelings of judgement compared to other humans.
Thus, it has become increasingly important for ECAs to build rapport and empathy to promote empathic grounding--an extension of grounding theory \cite{clark1991grounding}, in which the criterion for mutual understanding acknowledges the speaker's affective state, not just their spoken content \cite{arjmand2024empathic}.
Research shows that active listening, attending to, and acknowledging users using verbal and nonverbal cues signals attentiveness, empathy, and fosters social relationships with ECAs and encourages a willingness to engage openly in future conversations \cite{roos2023feeling,jang2024minimal, kobuki2023robotic}.

 \begin{figure}[t]
     \centering
\includegraphics[width=\linewidth]{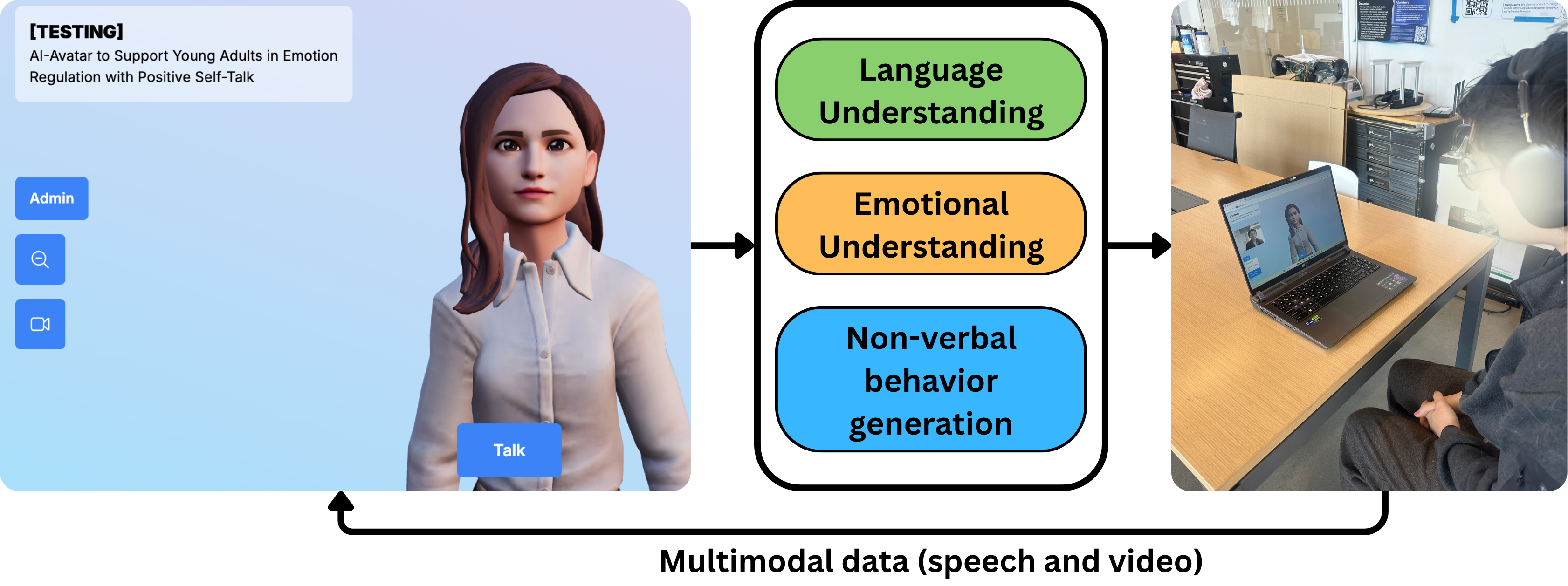}
     \caption{A multimodal AI-avatar system for backchanneling and emotional conversation support.}
     \Description{Three-panel diagram of TANDE's pipeline. Left: the ECA interface showing a female avatar with a Talk button and testing label. Center: three processing stages, Language Understanding, Emotional Understanding, and Non-verbal Behavior Generation, connected in sequence. Right: a photo of the physical study setup, a laptop displaying the avatar on a desk, with a participant seated nearby. An arrow labeled 'Multimodal data (speech and video)' loops from the study setup back to the interface, indicating the closed feedback loop.}
     \label{fig:tande}
 \end{figure}

One form of active listening is backchanneling.
Through verbal cues such as \textit{mhm}, \textit{yeah}, and \textit{uh-huh}; and nonverbal cues such as head nods, gestures, and gaze \cite{bavelas2000listeners, blomsma2024backchannel, duncan1972some}, backchanneling involves moment-by-moment exchanges of a listener actively attending to and acknowledging the speaker \cite{yngve1970getting, ward2006non}.
Otherwise, speakers lose confidence, and a sense of common ground and connection dissolves or is never created \cite{bavelas2000listeners}.
Furthermore, the use of \textit{empathic grounding} fosters social support and engagement, resulting in higher ratings of empathy and working alliance than an affectively neutral backchannel, though the task remained wizard-of-oz controlled and fixed \cite{arjmand2024empathic}. We focus on nodding and mhm, established backchannel signals in human conversation research \cite{bavelas2000listeners}, to test whether modality itself, independent of grounding content, shapes user perceptions.
We explore ECA  backchanneling characteristics with young adults, an underexplored area of research.

Young adults, aged 18-25 (\cite{NationalInstituteMentalHealth}), are increasingly using ECAs for emotional support, demonstrating high engagement and acceptability of AI-based mental health chatbots \cite{, koulouri2022chatbots, feijoo2024exploring}. This population shows the highest prevalence of mental illness \cite{NationalInstituteMentalHealth,kraaz2022backchannels, kobuki2023robotic} and they produce substantially more backchannels compared to other age groups \cite{kraaz2022backchannels, kobuki2023robotic}. 
As high producers of backchannels, young adults may hold stronger expectations around listener responsiveness, making them well-positioned to detect subtle differences in ECA backchanneling quality.
However, as ECAs become more realistic and behaviorally complex, they risk triggering uncanny valley responses, where a near-human appearance or behavior produces discomfort rather than connection \cite{mori2012uncanny}. 
Moreover, the backchannel modality could shape this effect, as unnatural or a mistimed feedback signal could increase rather than reduce uncanny valley perception in emotionally sensitive interactions.

Prior work on young adults and ECAs largely focuses on young adults’ perceptions of avatar realism and communication \cite{tauseef2025older, mikhailova2024age}, how young adults relate to avatars \cite{blinka2008relationship}, and the use of avatars for young adults' depression \cite{pinto2013avatar}.
These works do not consider or examine backchanneling behaviors, leaving a gap in understanding how these cues function in ECA interactions with young adults, particularly in emotional contexts.

Broader linguistic research indicates that backchanneling is shaped by gendered expectations. Prior work points to differences in how backchanneling is interpreted, with men more likely to perceive it as signaling agreement or conversational control, whereas women tend to interpret it as an indicator of attentiveness and engagement \cite{stubbe1998you, mulac1998uh}. 
These perceptual differences are especially relevant in ECA contexts, where embodiment further affects user expectations. 
For instance, ECAs with more human-like characteristics are perceived as more uncanny by males compared to females \cite{bailey2022observing}.
Despite these advances, modality-specific effects remain insufficiently understood in realistic interaction settings.
While some ECAs dynamically select verbal and nonverbal backchannels based on context or personality \cite{schroder2011building}, no prior work has experimentally isolated backchannel modality as a controlled condition, leaving the independent contributions of verbal and nonverbal cues on rapport, empathy, and engagement underexplored \cite{arjmand2024empathic, kato2025real}.

To address these gaps, we developed TANDE, a Large Language Model (LLM) powered, multimodal ECA system with independent verbal and nonverbal backchanneling, real-time facial emotion detection for \textit{empathic grounding}, and an open-domain backbone enabling conversation (see \autoref{fig:tande}).  
TANDE is beneficial because the backchannel rate can be fine-tuned for a target population using verbal, nonverbal, or both cues.
TANDE backchannels at a fixed-rate for young adults using the CANDOR Corpus \cite{reece2023candor}.
We investigate how distinct ECA backchannel modalities influence young adult perceptions in a general user study and through the lens of gender. Our study aims to answer these \textbf{research questions}:
\begin{enumerate}
    \item[\textbf{RQ1:}] How do ECAs' nonverbal and verbal-and-nonverbal back\-chan\-neling cues affect young adults' perceived rapport, empathy, and engagement?
    \item[\textbf{RQ2:}] How do female vs. male young adults perceive ECAs ability to build rapport, empathy, and engagement?
    \end{enumerate}

We hypothesize that \textbf{(H1)} ECA nonverbal cues will prove competitive with the combined condition, given that head nods alone produced the highest socio-affective ratings in ECA research \cite{compensis2026using}. 
Second, we predict that \textbf{(H2)} both active backchannel conditions will yield higher ratings of rapport, empathy, and engagement than the control, given that backchanneling consistently outperforms no backchanneling in ECA and robotic listening systems \cite{jang2024minimal, kobuki2023robotic}.

Our work makes the following \textbf{contributions}: (1) evidence suggesting that in empathically grounded ECAs, backchannel modality does not significantly affect rapport, empathy, or engagement in young adults, suggesting that content quality may matter more than listening behavior; (2) insights into young adults’ perceptions of ECA backchanneling, revealing that individual differences, particularly gender, shape responses more than modality, motivating personalized rather than age-group-level designs; and (3) design implications for emotionally supportive ECAs for young adults, a population with high mental health needs and limited care access.

\section{Background}

\subsection{Backchannels in Human Conversations}
Backchannels are short verbal and nonverbal listener signals, including tokens such as \textit{mhm}, \textit{yeah}, and \textit{uh-huh} as well as head nods and gaze, that convey attention and engagement without interrupting the speaker \cite{yngve1970getting, duncan1972some, bavelas2000listeners, blomsma2024backchannel}. Their absence degrades conversation quality and speaker confidence \cite{bavelas2000listeners}. 
Production varies by age, rising from near-absent in young children through adolescence into adulthood \cite{dittmann1972developmental, kraaz2022backchannels}. 
Gender also plays a role: women treat backchannels as signals of attentiveness, while men interpret them as agreement or conversational control \cite{mulac1998uh, stubbe1998you}. 
Backchannel modality shapes listener perception, particularly in emotionally sensitive contexts. Theoretically, nods affiliate with a speakers’s affective stance while vocal continuers signal only alignment with the telling activity \cite{stivers2008stance}. Empirically, the findings are mixed. In therapeutic contexts, an example rather than a direct comparison, therapists combining verbal acknowledgments and nodding were rated most empathic, while those using only one type rated lower than neither \cite{battles2012impact}. ECA research similarly found head nods alone outperformed verbal continuers, though combined modalities also rated highly \cite{compensis2026using}.

\subsection{ECA Backchanneling with Young Adults}
Early ECA research showed that nonverbal backchannel feedback tied to prosodic and movement cues positively influenced rapport and motivation \cite{gratch2006virtual}. Data-driven models outperform hand-designed timing rules in perceived naturalness \cite{murray2022learning}, with real-time prediction now deployable in live ECA systems \cite{inoue-etal-2025-yeah, paierl2025continuous}. Backchannel behavior also shapes ECA perception. Frequency and head nod amplitude influence perceived personality \cite{blomsma2022backchannel}, combining nonverbal and verbal backchannels increases affective trust in social robots \cite{anzabi2023effect}. However, emotionally specific backchanneling differs significantly between humans and human-robot interaction, suggesting models trained on human data may not transfer directly to ECA contexts \cite{shahverdi2023emotionally}, and motivating empirical evaluation of backchannel modality effects rather than assuming human-study findings generalize. 

Existing research involving young adults does not account for backchanneling, leaving a gap in understanding how it shapes their perceptions of rapport, empathy, and engagement with ECAs. Related work has examined photorealistic augmented reality avatars for interpersonal communication \cite{tauseef2025older}, aging and realism as avatar characteristics \cite{mikhailova2024age}, and avatar-based interventions in which young adults with depressive symptoms practice communication with virtual healthcare providers \cite{pinto2013avatar}.

\begin{figure}[t]
    \centering
    \includegraphics[width=\linewidth]{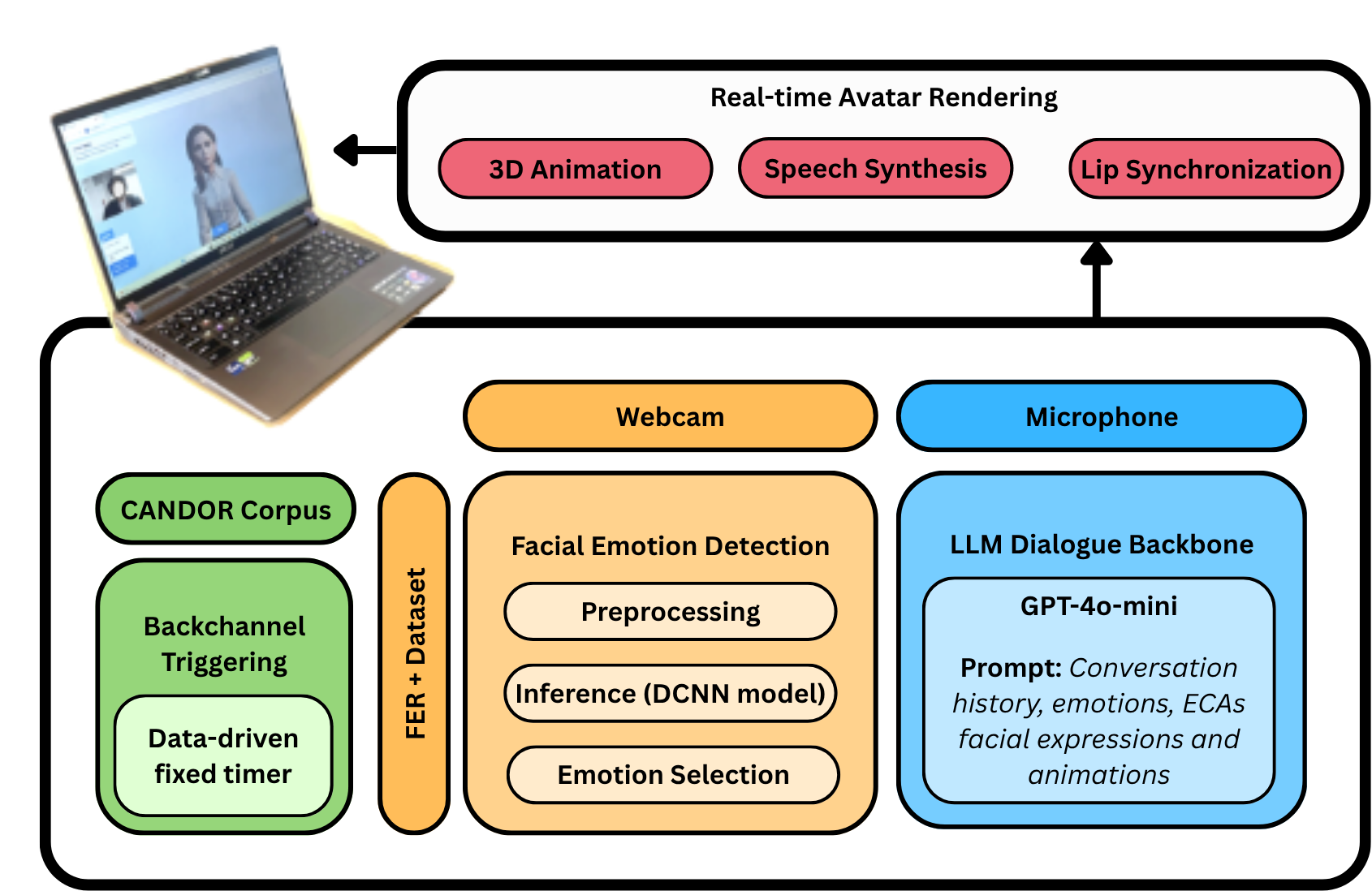}
    \caption{TANDE's architecture integrating facial emotion detection, LLM-based dialogue, and avatar rendering.}
    \label{fig:system-architecture}
    \Description{TANDE's system architecture. A laptop running the ECA interface sits at top-left. Below, an outer box represents the local processing pipeline which includes: Webcam feeds Facial Emotion Detection, which runs Preprocessing, Inference via a DCNN model, and Emotion Selection, using the FER+ Dataset. A Microphone feeds the LLM Dialogue Backbone, powered by GPT-4o-mini, prompted with conversation history, inferred emotions, and the ECA's facial expressions and animations. Separately, the CANDOR Corpus informs Backchannel Triggering via a data-driven fixed timer. All modules feed into Real-time Avatar Rendering at top, comprising 3D Animation, Speech Synthesis, and Lip Synchronization, which animates the avatar shown on the laptop screen.} 
\end{figure}

\subsection{Empathic Grounding in CAs}
Empathic grounding refers to conversational partners incrementally establishing mutual understanding, with listeners providing evidence of understanding through acknowledgments, demonstrations, and displays \cite{clark1991grounding}. Listeners play an active role in emotional conversations by providing affective responses that illustrate the emotional importance of the speaker’s narrative rather than merely signaling attention \cite{bavelas2000listeners}, motivating prior work on empathic listening ECAs capable of perceiving and producing affective response feedback \cite{devault2014simsensei, schroder2011building}. \citet{arjmand2024empathic} extends \citet{clark1991grounding}’s grounding theory through an empathic grounding system in which users’ facial expressions are continuously captured, smoothed in real-time, and integrated with speech before being processed by an LLM that generates multimodal grounding behaviors including affective facial expressions, head movements, and verbal utterances. This produced significantly higher ratings of empathy, emotional intelligence, trust, and working alliance than neutral backchannels, though the task remained wizard-of-oz controlled with limited flexibility \cite{arjmand2024empathic}. Prior work on digital therapeutic alliance further contextualizes these findings, showing that perceived empathy and practical support rather than trust alone are key drivers of long-term engagement with mental health chatbots \cite{xu2026chi}. Together, this suggests that affective backchanneling functions as a concrete mechanism through which conversational agents signal empathic attunement, motivating our investigation of its role in this work.

\section{TANDE: System Design and Implementation}
We introduce TANDE, inspired by the Kreyol word for “listen,” an LLM-driven ECA that 1) initiates questions, 2) receives user responses, and 3) iteratively perceives and analyzes users’ verbal and nonverbal behaviors to respond as a social support ECA (see Figure \ref{fig:system-architecture}).
Prior empathic grounding work has shown that automatically generated, affect-sensitive grounding moves can improve perceptions of empathy, emotional intelligence, trust, and working alliance, but this work was evaluated in a Wizard-of-Oz testbed in which a human operator controlled the task-level dialogue flow \cite{arjmand2024empathic}. TANDE extends this by implementing empathic grounding autonomously and using the system to manipulate backchannel modality while holding grounding, dialogue, ECA identity, and task framing constant across backchannel conditions. Thus, TANDE functions not only as an ECA, but also as a controlled experimental platform for examining the relative contributions of nonverbal and verbal backchannel cues.
Using data from CANDOR \cite{reece2023candor}, a fixed backchanneling trigger calibrated for young adults separates backchannel events from the main conversational floor, differing from prior approaches that rely on lexical matching or treat brief acknowledgments as full speaker turns, ensuring the resulting controller reflects listener behavior rather than turn-taking frequency.

TANDE is composed of four modules: 1) emotion detection from user facial expressions, 2) LLM-based dialogue backbone supporting unconstrained topic progression, 3) backchannel controller that triggers and independently activates verbal tokens and head nods, and 4) ECA rendering of real-time verbal and nonverbal responses.

\subsection{Implementing Empathic Grounding}
\label{sec:face}
To detect user emotions in real time, we implemented a privacy-preserving facial expression recognition pipeline through webcam video using the FER+ model \cite{barsoum2016training}, a VGG13-based deep convolutional neural network (DCNN) trained on soft label distributions from ten annotators per image rather than majority-vote labels, which improves generalization and expressiveness in unconstrained settings. FER+ provides eight emotion classes (\textit{neutral}, \textit{happiness}, \textit{surprise}, \textit{sadness}, \textit{anger}, \textit{disgust}, \textit{fear}, and \textit{contempt}) supporting our backchanneling logic and empathic grounding, where users' affective state is validated, ensuring TANDE’s responses are emotionally informed.

The model is deployed using ONNX Runtime Web\footnote{https://onnxruntime.ai/} with a WebAssembly\footnote{https://webassembly.org/} backend, enabling in-browser inference without transmitting video to a server, preserving privacy and maintaining low latency. Webcam frames are captured at approximately 5 frames per second and downsampled to 64$\times$64. When available, \texttt{OpenCV.js}\footnote{https://opencv.org/} applies Haar cascade face detection to crop and center the face region; otherwise, the full frame is used. RGB input is converted to grayscale using standard luminance weighting. The preprocessed frame is formatted as a $1 \times 1 \times 64 \times 64$ tensor, and passed to the FER+ ONNX model, which outputs softmax-normalized probabilities over the eight emotion classes. The highest-probability emotion is passed to the backchannel controller to inform real-time behavior, avoiding network dependencies critical in trust-sensitive applications.

The resulting emotion labels serve as a primary multimodal input to the dialogue controller, enabling affect-sensitive ECA responses rooted in real-time facial expressions captured by the DCNN. Emotion estimates are computed locally in the browser and passed to both the LLM and backchannel controller in real time, informing the timing and content of subsequent backchannels. Crucially, the affect-sensing module remains consistent across all experimental conditions, with emotion estimates computed locally and passed to the dialogue backbone and backchannel controller in real time, isolating observed effects to the backchannel modality rather than from changes in the emotion-recognition pipeline.

\subsection{LLM Dialogue Backbone}
\label{sec:llmbackbone}

TANDE uses \texttt{GPT-4o-mini}
\footnote{https://developers.openai.com/api/docs/models/gpt-4o-mini} as the underlying LLM because of its conversational performance, low latency suited to real-time dialogue, and cost efficiency. 
The model is prompted with the recent conversation history, the current inferred emotional state from facial analysis through FER+, and a response directive using an engineered prompt that signals empathic intent and encourages appropriate reflection and acknowledgment. 
This directive rewrites study task instructions (see \hyperref[sec:procedure]{Section~\ref{sec:procedure}}) into personalized ECA instructions. It constrains the model to a warm, non-judgmental tone, limiting responses to three messages per turn, and positioning TANDE as a reflective listener that guides users through one question at a time. 
As this was a controlled, non-clinical study involving young adults discussing general life stressors, prompt-level constraints on tone, role, and response scope were deemed sufficient guardrails for the experimental context. 
Future deployments beyond controlled experimental contexts should incorporate explicit safeguards for crisis-related disclosures, including detection and referral protocols for self-harm or suicidal ideation.

Unlike prior ECA empathic grounding research, where an LLM generated the agent's immediate response after each user turn but a human operator controlled the broader task dialogue, TANDE uses the LLM to support the autonomous conversational loop itself.
Given recent conversation history, the current inferred emotional state, and task-specific instructions, the model generates the ECA's next response without Wizard-of-Oz control over dialogue progression. This design keeps the dialogue system constant across conditions while backchannel modality is manipulated, allowing us to examine whether embodied listening cues provide perceived value beyond affect-sensitive conversational content. To support this independent loop, turn-taking is managed via a \texttt{TALK} button, which users click to speak; clicking the button again, now displayed in red as \texttt{STOP}, ends their turn and triggers TANDE’s response.

\subsection{Backchannel Triggering}
\label{sec:backchan}
The Backchannel Triggering module infers users’ real-time emotions via FER+ and \texttt{GPT-4o-mini}, using the full conversation history, and outputs a rate-based backchannel controller calibrated on naturalistic conversational data. This controller is con\-tent-inde\-pen\-dent by design, using no semantic content to decide when a backchannel fires, avoiding a possible confound between modality and content-driven timing. Emotion detection, conversation history, and LLM context shape response content, not backchannel timing.
Unlike prior empathic grounding work comparing affectively neutral backchannels with richer empathic behaviors, TANDE isolates the backchannel modality itself by separating the backchannel controller from the dialogue backbone, enabling verbal and nonverbal channels independently while the underlying conversation stays constant. This supports a controlled comparison of whether nonverbal or combined verbal-and-nonverbal backchanneling differentially shapes rapport, empathy, and engagement.
 
\paragraph{\textbf{Backchanneling Rate Derivation}}
We define $\mathrm{BPM}_{i,j}^{c}$ as a measure of backchanneling per minute of partner floor time and the total duration, in minutes, of all turns spoken by the conversational partner. Let $i \in I$ denote a focal speaker from the set of young adults, $j \in J$ denote the conversational partner of $i$, and $c \in C$ denote a conversation between $i$ and $j$. For each tuple $(i,j,c)$, let $b_{i,j}^{c}$ denote the total number of backchannel instances produced by speaker $i$ during conversation $c$, and $t_{j}^{c}$ denote the total partner floor time of speaker $j$ in that conversation. We then define backchannel rate as the number of backchannels produced by the focal speaker normalized by the amount of time the partner is holding the conversational floor. This formulation captures listener feedback behavior while controlling for variation in conversational opportunity.

\begin{equation}
\mathrm{BPM}_{i,j}^{c}
=
\frac{b_{i,j}^{c}}{t_{j}^{c}}
\end{equation}

\paragraph{\textbf{Dataset}} To quantify backchannel frequency with young adults, we used the CANDOR Corpus \cite{reece2023candor}, a large-scale dataset of 1656 English-speaking conversations with participants aged 19--66 ($M=33.3$, $STD=11.1$), including 788 females and 617 males. We adopted the Backbiter transcription variant, which preserves utterance-level backchannel events with timing and speaker metadata, allowing our  $BPM_{i,j}^{c}$ formulation to treat backchannels as annotated events using \texttt{utt.speaker.id} and \texttt{conversation\_id} to reconstruct speaker-conversation structure, \texttt{backchannel\_speaker} and \texttt{backchannel\_count} to attribute events, and \texttt{delta} to measure partner floor time, with \texttt{start}/\texttt{stop} as fallback for turn duration.

Recent studies show that people prefer ECAs that resemble their own behaviors in terms of demographics, particularly in mental health conversations \cite{feijoo2024exploring, feijoo2023participatory}. Despite these findings, we found minimal difference in backchanneling rates across six age groups in the CANDOR Corpus, revealing a tension in empirical data and embodied conversational agent literature (see supplementary materials).

\paragraph{\textbf{Preprocessing \& Implementation Details.}}
We employed four preprocessing techniques to compute $BPM_{i,j}^{c}$.
First, we filtered out conversations that did not involve two participants, including a young adult (aged 18 to 25, M$=$8.2, STD$=$3.3), catering to our target population and others, outside the target population. Second, we excluded observations with fewer than 3 minutes of total conversation time or fewer than 1 minute of partner floor time to reduce instability in short interactions, producing 977 dyadic conversations.
The backchannel trigger controller uses browser speech-recognition and sound events as a proxy for user speech activity. At the start of a listening session, the first possible backchannel is delayed by a uniformly sampled interval between 2.5 and 5.0 seconds. Later backchannels are scheduled from a target rate of $B_{i,j}^{c} = 8.164357$ events per minute, yielding an average gap of approximately 7.35 seconds. Each scheduled gap is perturbed by uniform \(\pm 35\%\) jitter, where ''jitter'' denotes a bounded random perturbation applied to an otherwise regular timing interval in order to avoid deterministic periodicity. The resulting gap is then clamped between 3.5 and 12.0 seconds, where ''clamping'' denotes constraining a computed value to a predefined admissible range in order to prevent extreme timing values. The controller increments its internal speech-time counter only while recent speech activity has been observed; specifically, speech activity remains active for up to 1.4 seconds after the most recent speech or sound event. Thus, backchannel triggers are constrained to periods in which the user is detected, or has very recently been detected, as speaking. 

Although real-time adaptive backchannel predictors have been developed using multimodal cues such as gaze, prosody, pauses, and lexical features~\cite{morency2008predicting,lala2022backchannel}, our goal was not to optimize backchannel timing adaptively. Instead, we use a rate-calibrated schedule with bounded jitter and clamping as an engineering compromise, preserving experimental control across conditions while introducing natural timing variability.
This timing design intentionally avoids two extremes: it is neither a deterministic metronome-like listener nor a fully adaptive model with unpredictable cross-participant behavior. Bounded jitter and clamping allow TANDE to approximate natural timing variability while preserving enough regularity for controlled comparison across conditions.

\subsection{Avatar Rendering and Speech Synthesis}
TANDE renders verbal and nonverbal responses using recognized emotions (see \hyperref[sec:face]{Section~\ref{sec:face}}), LLM outputs (see \hyperref[sec:llmbackbone]{Section~\ref{sec:llmbackbone}}), and backchannel triggers (see \hyperref[sec:backchan]{Section~\ref{sec:backchan}}).
While prior work relies on Wizard-of-Oz control to mitigate turn-taking errors \cite{arjmand2024empathic, weiss2015evaluating}, TANDE's facial expressions and animations for each turn are specified directly in the LLM's structured JSON response, enabling automatic affect-sensitive rendering rooted in conversational context. 

This structured rendering pipeline supports the experimental design by allowing the same underlying conversational system to be paired with different listening behaviors. Rather than treating backchannel behavior as a bundled ECA response, TANDE exposes verbal continuers and nonverbal nods as condition-specific components. This enables us to test whether adding verbal backchannels provides measurable benefits beyond nonverbal feedback alone.

The expressive behavior in the nonverbal and combined conditions is driven by a library of over ten FBX animation clips covering talking, idle, and emotional reactions, synchronized to a Web Audio API playback clock to ensure audio-visual alignment. 
TANDE's \textbf{speech synthesis} is generated through ElevenLabs TTS API\footnote{https://elevenlabs.io/text-to-speech-api}, with OpenAI TTS\footnote{https://developers.openai.com/api/docs/guides/text-to-speech} as a backup in case of API failures.
TANDE handles \textbf{lip synchronization} using Rhubarb\footnote{https://github.com/DanielSWolf/rhubarb-lip-sync}, which analyzes audio to produce a phoneme-to-viseme JSON mapping that drives mouth shape animations in real time;
\textbf{3D Animation} is controlled through a Wolf3D GLB rendered in the browser with morph targets controlling facial expressions and visemes.

\section{Experiment: Young Adult Case Study}
Our study targets the unique challenges young adults face. Based on the study tasks, potential topics of discussion could include future uncertainty \cite{schweizer2023uncertainty}, and socioeconomic and sociopolitical stressors \cite{Weissbourd_Batanova_McIntyre_Torres_Irving_Eskander_Bhai_2023}. The three task types illustrated in \autoref{fig:procedure} were designed to naturally surface these conversations.
To address this problem, we conducted a case study with young adults to understand how multimodal cues affect users' perceptions of emotion regulation.
We conducted a within-subjects in-lab experiment with $N=36$ young adults to understand (i) how nonverbal and verbal backchanneling affects perceived empathy, rapport, and engagement in emotional conversations with ECAs, and (ii) whether any backchanneling, compared to a static control, produces measurable differences.

\subsection{Study Design}
\begin{figure}[t]
     \centering
\includegraphics[width=\linewidth]{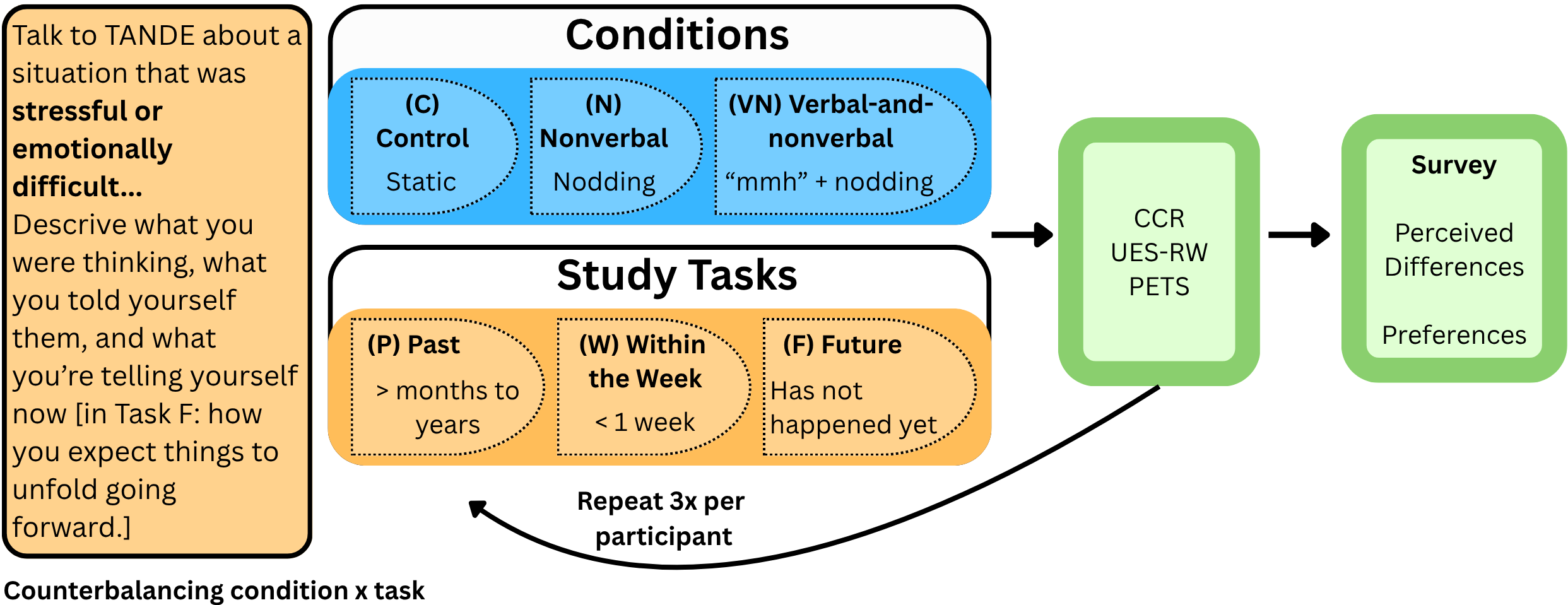}
     \caption{36 young adults interacted in a within-study design with the avatar, experiencing three conditions and tasks.}
     \label{fig:procedure}
    \Description{Study design diagram showing the experiments stages. Left: the task prompt shown to participants, asking them to describe a stressful or emotionally difficult situation, including what they were thinking then and are telling themselves now. Top center: the three experimental conditions, Control (static), Nonverbal (nodding), and Verbal-and-nonverbal ('mmh' plus nodding). Bottom center: the three study tasks by timeframe, Past (months to years ago), Within the Week (under 1 week ago), and Future (not yet happened). An arrow shows this condition-and-task pairing feeds into the CCR, UES-RW, and PETS survey measures, followed by questions on perceived differences and preferences. A looped arrow indicates the full sequence repeats three times per participant, with counterbalancing across condition and task.} 
 \end{figure}

The study followed a within-subjects 3×3 factorial design, with three conditions---control (C), nonverbal (N), verbal-and-nonverbal (VN)---and three tasks guiding user and AI conversations about stressful or emotionally difficult situations from the distant past (P), the last week (W), or the future (F). Counterbalancing with nine sequences was implemented using a Latin square design rather than full randomization to systematically control for order effects while ensuring that all condition–task combinations were evenly distributed across participants (see \autoref{fig:procedure} for details).
This study design, to our knowledge, is novel in isolating backchannel modality in live open-ended emotional conversation with an LLM-powered ECA, offering practical design insights for developers deciding whether to implement both verbal and nonverbal backchannel channels or whether one alone is sufficient for empathic grounding. This moves beyond the majority of prior work which compares backchanneling versus no backchanneling. A within-subjects design maximizes statistical power and minimizes individual differences as noise, ensuring observed effects reflect backchannel modality rather than participant characteristics.

\subsection{Participants}
We conducted an IRB-approved, in-lab experiment with $N = 36$ young adults between the ages of 18 and 25, with a mean age of $23$ years ($SD = 1.6$). Participants were recruited through word of mouth and flyers, and were compensated \$15. Of the 36 participants, 19 identified as female, 16 as male, and 1 preferred not to say. In terms of race, 24 identified as Asian, 11 as White, 2 as Black or African American, 1 as Middle Eastern or North African, and 1 preferred not to answer. 
Participants had to be proficient in everyday use of English and able to engage in spontaneous conversation and complex tasks. Participants were excluded if they had any cognitive or communication disorders that may have impaired these abilities. Twenty participants reported English as their first language; the remaining 16 participants reported their first languages as Spanish, Catalan, Mandarin, Hindi, or Italian. 

To determine the effect of our measures, we conducted a priori power analysis using G*Power \cite{faul2007g, faul2009statistical} F-test specifically for repeated measures ANOVA (within factors, 3 conditions). Effect size was estimated from pilot data of 10 participants using our primary outcome measure of rapport \cite{lin2025connection}. Partial eta square ($\eta^2 = 0.0254$) was converted to Cohen's $f = 0.1615$, with an average correlation of $0.7975$ between conditions, and the standard inputs of $\alpha = 0.05$ and a power of $0.80$. G*Power's results indicate that a required sample size of $N = 22$ was needed with an actual power of $0.8123$; we round this up to $N = 36$ for our full counterbalanced design.

\subsection{Procedure}
\label{sec:procedure}
Participants provided informed consent before completing a demographic questionnaire and brief study overview. They then interacted with TANDE in a private room for 5--8 minutes per session on a given laptop alongside a Qualtrics survey for post-surveys. Prior to each session, a researcher verified participants conditional group: control (C) condition, no nods or \textit{mhm} static ECA and only stands idle, speaks and gazes at participants; nonverbal-only (N) condition, nods, speaks, and does talking gestures; or combined verbal-and-nonverbal (VN) condition, everything in N with additional \textit{mhm}. Next, a researcher read one of three task descriptions of a stressful, anxiety-inducing, or emotionally difficult situation that was (P) months or years in the past, (W) within the last week, or (F) that they anticipated having in the near future  (See \autoref{fig:procedure}).

\subsection{Measures}
To investigate the effect of TANDE on participants' perceptions of backchanneling using nonverbal and multimodal cues, we collected a series of self-report measures, including 1) \textbf{rapport} using Connection-Coordination Rapport Scale (CCR), 2) \textbf{engagement}, through the Reward Factor subscale of the User Engagement Scale (RW-UES) \cite{o2018practical}, and 3) \textbf{perceived empathy}, through the Perceived Empathy of Technology Scale (PETS) \cite{schmidmaier2024perceived}. 
To protect participant privacy, we did not record audio data during interactions with TANDE due to potential sensitive topics that could arise, and to encourage participants to engage in open and honest conversations.

\subsection{Data Analysis}
We assess normality for all measures using a Shapiro-Wilk Test. 
CCR and PETS met the assumption for data normality, therefore required the use of ANOVAS, while RW-UES did not meet the assumption for normality, therefore required the use of a Friedman's ANOVA. To test our hypotheses and \textbf{RQ1}, two one-way repeated measures ANOVAs were conducted on CCR and PETS with condition (C, N, VN) as the within-subjects factor. Furthermore, to answer \textbf{RQ2}, we utilized participants' gender as a between-subjects factor using CCR, PETS, and RW-UES using a mixed-design ANOVA. Note that one participant was excluded as they were the only one in their gender group. Study tasks were excluded as a factor, as pilot data from the power analysis indicated similar CCR means across task conditions (Past: $M = 3.40$, $SD = 0.86$; Week: $M = 3.56$, $SD = 0.85$; Future: $M = 3.47$, $SD = 0.78$), and the counterbalanced design ensured orthogonality.
\begin{table}[t]
\caption{Mixed-design ANOVA statistics are reported for between-subjects effects of gender. Statistically significant results are marked with asterisks (* $p < .05$, ** $p < .01$)}
\small
\begin{tabular}{llll}
\hline
Factor & CCR & PETS & RW-UES \\
\hline
Condition & $F(1, 33) = 5.449$ & $F(1, 33) = 8.163$ & $F(1, 33) = 1.128$ \\
          & $p = .026^{*}$         & $p = .007^{**}$         & $p = .296$ \\
          & $\eta_p^2 = .142 $ & $\eta_p^2 = .198$ & $\eta_p^2 = .033$ \\
\hline
\end{tabular}
\label{tab:gender-means}
\end{table}

\begin{figure}[t]
\centering
\includegraphics[width=\linewidth]{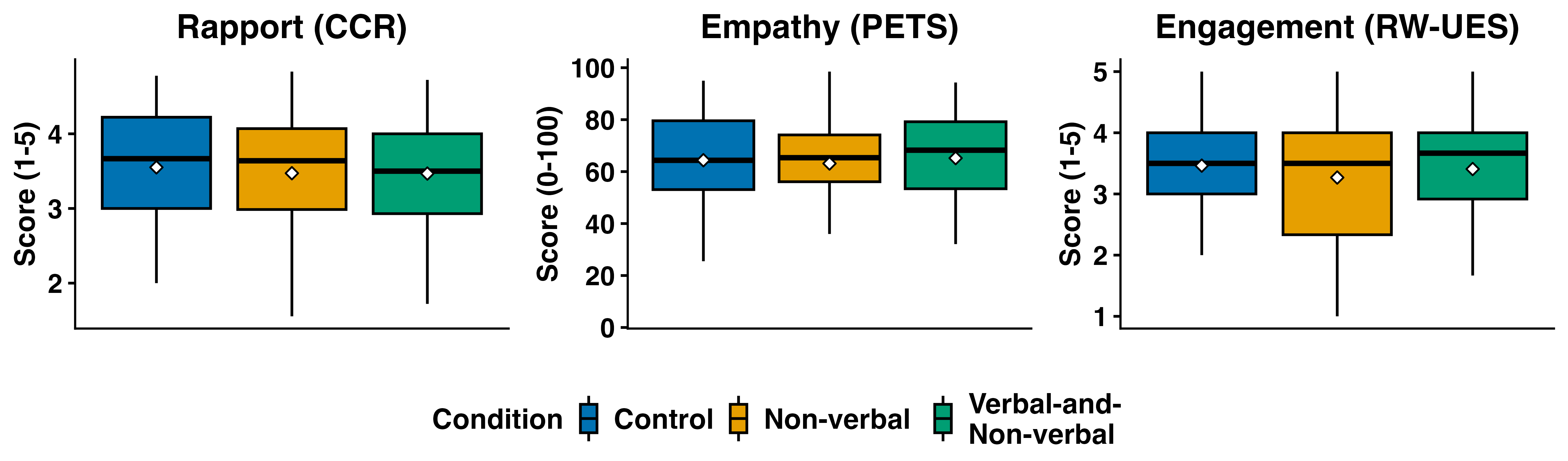}
\caption{Rapport (CCR), Empathy (PETS), and Engagement (RW-UES) ratings across Control, Nonverbal, and Verbal and Nonverbal conditions. Diamonds indicate the mean for each condition.}
\label{fig:primary-result}
\Description{Three box plots comparing Rapport (CCR, scored 1–5), Empathy (PETS, scored 0–100), and Engagement (RW-UES, scored 1–5) across the three conditions: Control (blue), Nonverbal (orange), and Verbal-and-Nonverbal (green). Diamond markers show the mean for each condition. Across all three measures, the boxes largely overlap between conditions, with no condition showing a clearly higher distribution than the others.}
\end{figure}

\begin{figure}[t]
\centering
\includegraphics[width=\linewidth]{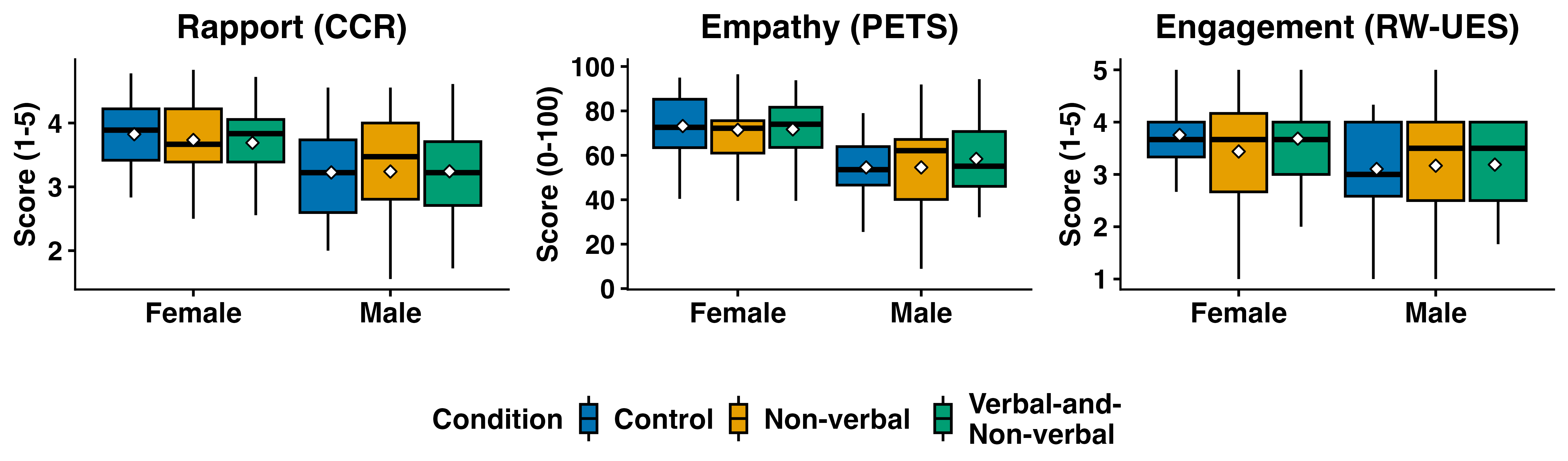}
\caption{Gender distribution of Rapport (CCR), Empathy (PETS), and Engagement (RW-UES) ratings across Control, Nonverbal, and Verbal and Nonverbal conditions. Diamonds indicate the mean for each condition.}
\label{fig:gender-means}
\Description{Six box plots showing the same three measures, Rapport (CCR), Empathy (PETS), and Engagement (RW-UES), split by gender (Female, Male) and by condition (Control, Nonverbal, Verbal-and-Nonverbal) within each gender group. Female participants' boxes sit visibly higher than male participants' across most conditions for Rapport and Empathy, consistent with the reported significant gender main effects; Engagement shows more overlap between genders.}
\end{figure}

\section{Quantitative Results}

\paragraph{\textbf{Gender Rapport (CCR), Engagement (RW-UES), Empathy (PETS)}}

\autoref{fig:gender-means} and \autoref{tab:gender-means} show the mean scores for the between-subjects factor of gender showed that female participants consistently reported higher levels of CCR, PETS, and RW-UES across all three experimental conditions compared to male participants, addressing \textbf{RQ2}. Specifically, there were significant main effects of gender for Rapport ($F(1,33) = 5.449, p = .026$) and Empathy ($F(1,33) = 8.163, p = .007$). Due to a violation of the assumption of normality for the engagement (RW-UES) data, a non-parametric Aligned Rank Transform \cite{wobbrock2011aligned} was conducted, yielding a non-significant result ($F(1,33) = 1.128, p = .296$). 
No significant main effect of Condition was found for CCR ($F(2, 66) = 0.24, p = .789$), PETS ($F(2, 66) = 0.66, p = .520$), and RW-UES ($F(2, 66) = 0.22, p = .802$). Furthermore, the interaction between Condition and Gender was not significant for CCR ($p = .668$), PETS ($p = .313$), and RW-UES ($p = .558$)

\paragraph{\textbf{Rapport (CCR), Empathy (PETS), Engagement (RW-UES)}}
Mean scores were similar across conditions for all three measures, thereby not supporting \textbf{H1} or \textbf{H2}. The control condition received slightly higher mean scores on CCR (rapport) and RW-UES (engagement), while verbal-and-nonverbal received slightly higher mean scores on PETS (empathy); see \autoref{fig:primary-result}. 
Though a one-way repeated measures ANOVA found no significant effect of Condition on CCR, $F(2,70) = 0.586$, $p = .559$, nor on empathy, $F(2,70) = 0.722$, $p = .489$, and  Friedman's ANOVA for engagement showed no significant effect, $\chi^2_F(2) = 2.667$, $p = .264$. 

\paragraph{\textbf{ECA Preferences}} $38.9\%$ of participants ($n = 14$) preferred N, $33.3\%$ ($n = 12$) preferred C, $22.2\%$ ($n = 8$) preferred VN, and $5.6\%$ ($n = 2$) did not notice any differences. 

\section{Qualitative Results}

Participants’ open-ended responses offered additional insight into their perceptions of TANDE’s backchannel behaviors, independent of their survey ratings. We identified three themes:

\paragraph{\textbf{Mixed reactions to verbal backchanneling}}
Participants had mixed reactions to the ECA’s \textit{mhm} sounds. Thirteen participants stated they appreciated the verbal feedback, describing it as a clear signal of attentiveness. P15 noted that the verbal cues made them ``feel like it is trustful and the ECA is actually listening,’’ and P34 similarly explained that the verbal condition ``would make noises to ensure that I knew she was listening.’’ One participant added that ``the nodding felt natural and the mhm was reassuring...without them the conversation did feel more dull.’’

However, 15 participants had either mixed views or outright disliked the \textit{mhm} cues reported that it was due to a lack of naturalness, timing, and frequent interruptions. P32 stated that ``it felt like I was constantly getting cut off.’’ Some even found the verbal backchanneling triggered uncanny valley effects. P17 commented that ``when the avatar gave audible listening feedback it felt more synthetic since I knew this wasn’t honest feedback,’’ while P19 reported they did not expect an ECA to \textit{mhm} in conversations and that it felt too close to human behavior and that it interrupts their thoughts. P21 also noted that the volume of the \textit{mhm} was too loud; in human conversations, they expected someone’s \textit{mhm} to be quieter than their speaking voice, rather than at the same level.

\paragraph{\textbf{Nodding as a signal of attentiveness}}
In contrast to the polarized views on verbal backchannels, nonverbal backchanneling (nodding) was consistently described positively by 25 of the participants. P9 noted that the ECA ``nodded during key emotional moments as I spoke (making me feel it was emotional), which was nice.’’ Other participants remarked that nodding was natural ``because that’s what humans usually do during conversations,’’ while another wrote that ``the nodding makes me feel more at ease.’’

\paragraph{\textbf{Static vs. Animated ECAs}} Fourteen participants stated that they liked when the ECA was still either exclusively or had some preference, finding they liked less movement compared to nonverbal movements and verbal cues. P21 wrote that the control ECA ``was completely inanimate, and I actually preferred that as it lets me focus on how I feel rather than a 3D animation.’’ P22 similarly reflected that ``the avatar animations felt very unnatural, which weirdly made the avatar feel more natural when it wasn’t animated.’’ Both P21 and P22 further stated that they did not want ECAs at all during these interactions. P4 further states:  ``I think the avatar who did neither was the least distracting as I was speaking, and I felt like I was being listened to. I feel that I kind of do the same as well when listening to someone, so I may be biased in that sense.''

\section{Discussion}
Our results showed that backchannel modality had no significant effect on rapport, empathy, or engagement \textbf{(RQ1)}. Female participants rated the ECA significantly higher on rapport and empathy than male participants; engagement did not differ significantly \textbf{(RQ2)}. For our first hypothesis, we predicted that N would perform as well as VN, but was not supported as nonverbal did not outperform the other conditions \textbf{(H1)}. Lastly, H2 predicted that both N and VN would outperform control, but this was not supported as the control scored high or slightly higher \textbf{(H2)}. 

\subsection{Low-Interruptive Signals of Attentiveness}
The results suggest that a majority of participants found nonverbal backchanneling sufficient, and that adding verbal cues provided no additional perceived benefit. This diverges from prior work showing that emotional support providers using both verbal acknowledgments and nodding were perceived as most empathic, while those using only one type were perceived as less empathic \cite{battles2012impact}. This prior finding comes from human therapists in clinical settings, whereas TANDE is a non-clinical ECA interacting with young adults. Thus, differences in agent, setting, and population may help explain this divergence.
In our study, however, the combined condition was the least preferred, while nonverbal and control were preferred more. This suggests the focus should remain on users’ reflection and self-disclosure in emotional conversations rather than the ECA drawing attention to itself through verbal feedback.

This is reflected in participants’ responses, which suggest they did not expect or want TANDE to behave like a human therapist, pointing to different expectations for non-clinical ECAs. Fifteen participants noted verbal cues felt artificial or too close to human behavior, while others preferred the static ECA because it allowed them to focus on their own reflection. These reactions are consistent with research showing that users apply distinct trust frameworks to machines versus humans and that increasing ECA’s human-likeness raises expectations in ways that can produce negative outcomes when unmet \cite{go2019humanizing, strait2015too}. Young adults navigating significant emotional and identity challenges may accept human-like conversational acknowledgments from an ECA, potentially finding that verbal cues interrupt, rather than support self-reflection.

\textbf{Design Recommendation (DR) 1.} ECAs need \textit{subtle} nonverbal backchanneling cues rather than verbal cues that may draw attention away from self-reflection in emotional conversations. 

\subsection{LLM-Generated Conversations}
Findings on participants' preferences do not explain why backchannel modality produced no significant differences in our quantitative measures of rapport, empathy, and engagement. One explanation is that the empathically grounded LLM responses were sufficiently meaningful to participants regardless of the backchannel condition. All three conditions shared the same underlying LLM content, differing only in what the ECA did behaviorally, and that content alone may have been the primary driver of perceived rapport and empathy. 
Because TANDE used an empathic grounding prompt that validated users’ emotional states in real time, participants may have experienced a sufficiently meaningful and supportive conversation regardless of whether the ECA nodded, verbalized, or remained static.
Furthermore, backchannel modality may matter more for ECAs that produce low-quality responses when affective grounding is absent, a concern we leave for future work.

\textbf{DR 2.} Researchers building ECAs should first focus on designing empathic grounding communication cues before investing effort in refining backchannel modality. When LLMs have context about users' emotional state, the added benefit of varying multimodal backchanneling cues may be negligible. 

\subsection{Impact of ECA Complexity on Emotional Support}
\textit{What is the appropriate level of ECA complexity to produce effective, emotional support?}
A substantial portion of participants ($N = 12$) actively preferred a static ECA with no backchanneling.
Instead, they placed more emphasis on the content of the conversation rather than TANDE's behavioral responses.
This finding is consistent with prior work that shows some users favor simpler LLM-based chatbots over complex interfaces \cite{tan2025exploring}. Further evidence demonstrates that animated ECA chatbots can produce more discomfort and physiological arousal than simpler text-based alternatives \cite{ciechanowski2019shades}. These findings together suggest that users prefer simpler ECA multimodal interactions, and increasing the complexity of such interactions has the potential to draw attention away from self-reflection in emotional conversations. While ECAs have well-documented benefits in other domains \cite{mercado2023embodied}, their need in emotion-focused reflective applications warrants further investigation. 

\textbf{DR 3.} Designers should critically evaluate whether ECA embodiment is necessary for their specific context rather than adopting it by default in emotional conversations. Simpler modalities may better support the user's focus on self-reflection, though further research is needed to understand these trade-offs long-term.

\subsection{Gender Differences in ECA Perception}
Our findings suggest that backchannel preferences vary substantially across individuals and genders, with no single condition preferred by a majority of participants. 
More specifically, we found gender was a significant predictor of rapport and empathy across conditions, with female young adults rating the ECA as building rapport and showcasing greater empathy compared to male young adults. There are many explanations for why our results show this trend. 
Prior research has found that people prefer ECAs that match their gender, for example, women prefer a female ECA, though men did not place importance on self-similarity \cite{probster2023like}. 
Furthermore, women interpret backchanneling as attentiveness and engagement \cite{mulac1998uh, stubbe1998you}, which are factors related to rapport and empathy \cite{buck1990rapport}.
Women also tend to express emotions more freely than men \cite{goldshmidt2000talking}, which may make women more attuned and appreciative of ECAs empathetic responses in our empathic grounding framework. However, this backchanneling interpretation does not fully account for our results, as female participants also reported higher social connectedness in the control condition, where no backchanneling occurred, pointing to broader gender differences in ECA perception.
Thus, perhaps TANDE's gender could have affected our results, where male participants would have potentially shown more favorable impressions towards a male version of TANDE. 
Future research should examine users’ preferences for ECA gender.

\textbf{DR 4.} ECA designers should allow users to personalize the ECAs they interact with to enhance user perceptions.

\subsection{Limitations and Future Work}

While our findings provide useful insights into how gender groups respond to ECA backchanneling, there are some limitations worth mentioning.
First, TANDE is a general-purpose ECA whose young-adult focus enters only through the backchannel rate and task descriptions. It can serve other populations by re-calibrating both, though our results are limited to young adults in short-term interactions.
Second, TANDE’s backchannel generation was intentionally simple rather than an advanced, adaptive model. This simplicity does not appear to have limited our contribution, as individual and gender differences, not modality, drove the observed effects. Similarly, because our backchannel trigger is rate-based rather than content-aware, verbal cues could fire at semantically inappropriate moments, which may explain why several participants described \textit{mhm} as unnatural or interruptive, independent of verbal backchanneling as a modality.
Third, participant interactions with TANDE were short-term, 7 minutes and 30 seconds on average; thus, our findings do not reflect users' perceptions of ECAs in long-term interactions. 
Fourth, we deployed TANDE as a female ECA to reduce the scope of our work; however, this design choice likely affected users' perceptions of interactions with TANDE.
Lastly, our quantitative results for (\textbf{RQ1}) showed no significant effect of backchannel modality on any outcome measure, which indicates a larger sample size could have produced higher effects on users' perceptions of TANDE. 
To address these limitations, we propose to conduct further studies, allowing participants to customize their ECA, expanding to other populations of people with social support needs in long-term interactions, and adaptive multimodal backchanneling capabilities to further investigate ECA design through a gender lens. 

\section{Conclusion}
In this work, we examined how young adults perceive backchanneling modalities (control, nonverbal-only, and verbal-and-nonverbal) in terms of rapport, engagement, empathy, and gender differences. With $N = 36$ young adults, we found no difference across conditions overall, but isolating for gender revealed female participants rated the ECA significantly higher in rapport and empathy, suggesting user characteristics may shape perceived relational quality more than backchannel design itself. Qualitative results revealed mixed reactions to the conditions, with a preference for nonverbal-only backchanneling, implying less intrusive cues may better support users’ self-reflection. From these results, we derived design recommendations on embodiment and conversational content for young adults’ emotional conversations with ECAs. This work addresses a gap in understanding backchanneling modalities for young adults, a population with high mental health needs and limited access to care, and raises open questions for the multimodal community on the use of ECAs in these contexts.

\section{Safe and Responsible Innovation Statement}
Ethical considerations for this work include IRB-consent before experimentation, informed consent, and participants’ rights to discontinue participation at any point during the study. All participants’ quantitative and qualitative data were anonymized and securely stored on university-approved servers. Our study provides insights into the roles of multimodal interactions of backchanneling and of ECAs in emotional conversations. 

\bibliographystyle{ACM-Reference-Format}
\bibliography{references}

\end{document}